\documentclass[12pt]{article}
\usepackage{amssymb}
\begin{document}
\title{\bf Classification of Cosmic Scale Factor via Noether Gauge Symmetries}
\author{Adil Jhangeer $^{(1)}$\thanks{adil.jhangeer@gmail.com},
M. Farasat Shamir $^{(2)}$\thanks{farasat.shamir@nu.edu.pk}, \\Tayyaba
Naz$^{(2)}$ \thanks{{tayyaba.naz42@yahoo.com}} and Nazish Iftikhar$^{(2)}$ \thanks{nazish.iftikhar289@gmail.com}.\\\\  $^{(1)}$Deanship of Educational Services, Preparatory Year Unit, \\Qassim University, \\P.O. Box 6595, Al-
Qassim, Buraidah: 51452 \\Kingdom of Saudi Arabia.\\ $^{(2)}$Department of
Sciences and Humanities, \\National University of Computer and
Emerging Sciences,\\ Lahore Campus, Pakistan.}

\date{}
\maketitle

\begin{abstract}
In this paper, a complete classification of
Friedmann-Robertson-Walker (FRW) spacetime by using approximate
Noether approach is presented. Considered spacetime is discussed for
three different types of universe i.e. flat, open and closed.
Different forms of cosmic scale factor $a$ with respect to the
nature of the universe, which posses the nontrivial Noether gauge
symmetries (NGS) are reported. The perturbed Lagrangian
corresponding to FRW metric in the Noether equation is used to get
Noether operators. For different types of universe minimal and
maximal set of Noether operators are reported. A list of Noether
operators are also computed which is not only independent from the
choice of the cosmic scale factor but also the choice type of
universe. Further, corresponding energy type first integral of
motions are also calculated.
\end{abstract}

{\bf Keywords:} Friedmann-Robertson-Walker, Noether gauge
symmetries, Scale factor.\\\\
{\bf PACS:} 04.50.Kd, 98.80.-k, 02.20.Sv.

\section{Introduction}
FRW models of the universe are known to
be spatially homogeneous and isotropic in nature. This indicates
that these models are the best representation of the large scale
structure of our present universe. The FRW spacetime is given by
\begin{equation}
ds^{2}=dt^{2}-a^2\bigg[\frac{dr^{2}}{1-kr^2}+r^{2}d\Omega^{2}\bigg],
\label{c1}
\end{equation}
where $d\Omega^{2}=(d\theta^2+\sin^2\theta d\phi^2)$ and the
curvature parameter $k$ is $0$, $1$ or $-1$, which represents flat,
open or closed universe respectively. Here $a$ is a function of
cosmic time $t$ and commonly known as scale factor of universe. It
is also called Robertson-Walker scale factor parameter of the
Friedmann equations which represents the expansion of the universe.
The importance of scale factor can be well understood from the facts
that it is used in defining some important terms like the Hubble
parameter and red shift. It has been established \cite{2}-\cite{3}
that the universe is now expanding with an accelerated rate which
means that the second derivative of the scale factor with respect to
cosmic time should be positive.
\par
Cosmological solutions in multidimensional model with multiple
exponential potential are obtained by Ivashchuk et al.
\cite{Ivashchuk}. They classified the solutions with power law and
exponential behavior of scale factors. Similar classification of
scale factor has been used by Sharif et al. \cite{14} to investigate
Bianchi Type $I$ and $V$ solutions in $f(R)$ theory of gravity. They
used the ansatz method to find the solutions and classification of
scale factor. It would be interesting to classify the scale factor
using Noether approach.

\par A first systematic approach for finding the conserved quantities \cite{CSC,
TTSP} of the variational problems is given by a German mathematician
Emmy Noether. She reported a relation between the symmetries and
conservation laws \cite{TTSP}. These symmetries are called Noether
symmetries. She concluded that each symmetry corresponds to a
conserved quantity. The idea of conservation laws in mathematics
came from devising the physical laws for conserved quantities
like mass, energy and momentum.

Conservation laws play a vital role in the field of differential
equations. They help to compute the unknown exponent in the
similarity solution which cannot be obtained from homogeneous
boundary conditions \cite{NMM1}. The conserved quantities are also
helpful to control numerical errors in the numerical integration of
partial differential equations (PDEs). It would be interesting to
mention here that many physical solutions of general relativity
field equations do possess some symmetry \cite{Bokhari and Kara}.
Bokhari et al. \cite{kashif} investigated the symmetries generated
by Killing vectors for some spacetimes. They established that
Noether symmetries of these spacetimes are more than the symmetries
generated by Killing vectors. Thus it was conjectured that the
Noether symmetries always provided a larger set of Lie algebra and
Killing symmetries formed a subalgebra of Noether symmetries.
\par

Modified theories of gravity have attracted much attention of the
researchers during the last decade. Jamil et al. investigated some
new exact solutions of static wormholes in the context of $f(T)$
gravity \cite{T24}. The same authors discussed the energy conditions
in generalized teleparallel gravity models \cite{T25}. In addition
to this, violation of first law of thermodynamics is also discussed
in $f(R, T)$ gravity \cite{T26}. The resolution of dark matter problem has
been studied in $f(T)$ gravity \cite{T27}. Bamba et a1. \cite{T30}
have investigated the generalized second law of thermodynamics with
entropy corrections in the field of $f(T)$ gravity. Similarly many
authors explored modified theories in different contexts
\cite{T21}-\cite{T23}. Noether symmetries have also been used to
deal many issues in modified theories of gravity. Kucukakca and
Camci \cite{T43} have obtained the function $f(R)$ and the scale
factor  in Palatini $f(R)$ theory via Noether symmetry approach.
Jamil et a1. \cite{T49} discussed the $f(R)$- Tachyon model in the
metric formalism and also resulted a zero gauge function. Bianchi
type $I$ cosmology in generalized Saez-Ballester theory has also been
studied by using the definition of NGS \cite{T44}. Hussain et al.
\cite{T45} found Noether symmetries for the flat FRW model using the
gauge term in metric $f(R)$ gravity.

Noether symmetries are helpful in recovering some lost conservation
laws and symmetry generators of spacetimes \cite{lost}. Sharif et
al. \cite{siara1} explored the energy contents of colliding plane
waves using approximate Noether symmetry approach. It has been
concluded that for plane electromagnetic and gravitational waves
there does not exist any non-trivial first order symmetry generator.
The same authors \cite{siara2} used approximate symmetries to
investigate the energy contents of Bardeen model and stringy charged
black hole solutions. Capozziello et al. \cite{A 20} investigated
$f(R)$ gravity for spherically symmetric spacetime using Noether
symmetry. In \cite{A 21}, a new class of Noether symmetries has been
reported for spherical symmetry in $f(R)$ gravity. Paliathanasis et
al. \cite{A 22} reviewed the modified $f(R)$ gravity models using
the Lie and Noether point operators. In a recent paper, Shamir et
al. \cite{shamir} investigated $f(R)$ gravity using Noether gauge
symmetries. For this purpose, Noether symmetry generators for FRW
universe and spherically symmetric spacetimes are evaluated. The
corresponding conserved quantities are also obtained along with
importance and stability criteria of some specific $f(R)$ models.
Here our attention is to discuss the cosmic scale factor $a$ using
NGS approach.
\par The paper is organized as
follows: Section $2$ is devoted for the fundamental operators, while
in Section $3$ classification of scale factor by Noether approach
for different types of universe is reported.

\section{Fundamental operators} Let
\begin{eqnarray}
ds^{2}= g_{ij} dx^{i}dx^{j}, \label{T1}
\end{eqnarray}
be a line element then the vector field $X$ for (\ref{T1}) will be
\begin{eqnarray}
X=\xi(s,x^{i}){\frac{\partial}{\partial{s}}}+\eta^{j}(s,x^{i}){\frac{\partial}{\partial{x^{j}}}}~.
\label{T2}
\end{eqnarray}
The Lagrangian $L$ for metric (\ref{T1}) is given by \cite{19,20}
\begin{eqnarray}
L= \frac{1}{2}g_{ij}{\dot{x}^{i}}{\dot{x}^{j}},\label{T3}
 \end{eqnarray}
where dot represent derivative with respect to $s$. The Noether
equation is:
\begin{eqnarray}
X^{[1]}(L)+ L D_{s}(\xi)=D_{s}(A), \label{T4}
\end{eqnarray}
where $A$ is a gauge function and $X^{[1]}$ is the first order
prolongation and $D_{s}$ is called total derivative defined as
\begin{eqnarray}
D_{s}={\frac{\partial}{\partial{s}}}+{\dot{x}^{i}}{\frac{\partial}{\partial{x}^{i}}}~.
\end{eqnarray}
The energy type first integral of motion also known as conserved
quantity corresponding to Noether operator $X$ is defined as
\begin{eqnarray}
I=\xi L+(\eta^{i}-\xi{\dot{x}^{i}}){\frac{\partial
L}{\partial{\dot{x}^{i}}}}-A.
 \label{c4}
\end{eqnarray}
\section{Energy contents for FRW metric}
The Lagrangian for FRW spacetimes (\ref{c1}) is
\begin{eqnarray}
L=
{\dot{t}}^{2}-\bigg(\frac{a^2{\dot{r}}^{2}}{1-kr^2}\bigg)-a^2r^2{\dot{\theta}}^{2}-a^2r^2{\dot{\phi}}^{2}\sin^{2}\theta.
 \label{c6}
 \end{eqnarray}
The corresponding vector field $X$ is defined as
\begin{eqnarray}
X=\xi(s,t,r,\theta,\phi){\frac{\partial}{\partial{s}}}+\eta^{1}(s,t,r,\theta,\phi){\frac{\partial}{\partial{t}}}+\eta^{2}(s,t,r,\theta,\phi){\frac{\partial}{\partial{r}}}+\nonumber
\\\eta^{3}(s,t,r,\theta,\phi){\frac{\partial}{\partial{\theta}}}+
\eta^{4}(s,t,r,\theta,\phi){\frac{\partial}{\partial{\phi}}}.
\nonumber
\end{eqnarray}
Next we will discuss three different cases with respect to the value
of $k$.
\subsection{For flat universe ($k= 0$)}
The Lagrangian (\ref{c6}) for this case becomes
\begin{eqnarray}
L={\dot{t}}^{2}-a^2{\dot{r}^{2}}-a^2r^2{\dot{\theta}}^{2}-a^2r^2{\dot{\phi}}^{2}\sin^{2}\theta.
\label{c7}
\end{eqnarray}
Substituting Eq.~(\ref{c7}) in Eq.~(\ref{T4}) and after some
manipulation we get an over-determined system of linear PDEs
\begin{eqnarray}
(i)~\xi_{\phi}=0,~~~~~(ii)~\xi_{\theta}=0,~~~~~(iii)~\xi_{r}=0,~~~~~
(iv)~\xi_{t}=0, ~~~~~ (v)~A_{s}=0,~\label{c8}
\end{eqnarray}
\begin{eqnarray}
(i)~2\eta^{1}_{s}=A_{t},~~~~~
(ii)~-2a^{2}\eta^{2}_{s}=A_{r},~~~~~(iii)~-2a^{2}r^{2}\eta^{3}_{s}=A_{\theta},~~
\label{c9}
\end{eqnarray}
\begin{eqnarray}
(i)~-2a^{2}r^{2}\sin^{2}\theta\eta^{4}_{s}=A_{\phi},~~~~~(ii)~\eta^{1}_{r}-a^{2}\eta^{2}_{t}=0,
\label{c10}
\end{eqnarray}
\begin{eqnarray}
(i)~\eta^{1}_{\theta}-a^{2}r^{2}\eta^{3}_{t}=0,~~~~~(ii)~\eta^{1}_{\phi}-a^{2}r^{2}\sin^{2}\theta\eta^{4}_{t}=0,
\label{c11}
\end{eqnarray}
\begin{eqnarray}
(i)~a^{2}r^{2}\eta^{3}_{r}~+~a^{2}\eta^{2}_{\theta}=0,~~~~~(ii)~a^{2}\eta^{2}_{\phi}+a^{2}r^{2}\sin^{2}\theta\eta^{4}_{r}=0,
\label{c12}
\end{eqnarray}
\begin{eqnarray}
(i)~a^{2}r^{2}\eta^{3}_{\phi}+a^{2}r^{2}\sin^{2}\theta\eta^{4}_{\theta}=0,~~~~~(ii)~2\eta^{1}_{t}-\xi_{s}=0,
\label{c13}
\end{eqnarray}
\begin{eqnarray}
a^{2}\xi_{s}-2a^{2}\eta^{2}_{r}-2aa_{t}\eta^{1}=0, \label{c14}
\end{eqnarray}
\begin{eqnarray}
-2a^{2}r\eta^{2}+~a^{2}r^{2}\xi_{s}-2a^{2}r^{2}\eta^{3}_{\theta}-2aa_{t}r^{2}\eta^{1}=0,
\label{c15}
\end{eqnarray}
\begin{eqnarray}
a^{2}r^{2}\sin^{2}\theta\xi_{s}-2a^{2}r^{2}\sin^{2}\theta\eta^{4}_{\phi}-2a^{2}r\eta^{2}\sin^{2}\theta-
2a^{2}r^{2}\sin\theta\cos\theta\eta^{3}-\nonumber
\\2aa_{t}r^{2}\sin^{2}\theta\eta^{1}=0.
\label{c16}
\end{eqnarray}
\subsubsection{Minimal set of Noether operators}
 If $a(t)$ is arbitrary then one can get minimal number of Noether symmetries:
 $$\bf{X_{01}}=\frac{\partial}{\partial{s}},~~\bf{X_{02}}=\frac{\partial}{\partial{\phi}}$$
 where the gauge function $A$ is constant.
The corresponding energy type first integral of motions associated
with $\bf{X_{01}}$ and $\bf{X_{02}}$ are:
$$\bf{I_{01}}=-{\dot{t}}^{2}+a^2{\dot{r}^{2}}+a^2r^2\dot{\theta}+a^2r^2\sin^{2}\theta{\dot{\phi}}^{2},~~~~
I_{02}=-2a^2r^2\sin^{2}\theta{\dot{\phi}}.$$

\subsubsection{Five Noether operators}
Some different forms of $a(t)$ which posses five Noether operators
are listed in Table: $1$ and the set of Noether symmetries consist
of $\bf{X_{01}}$, $\bf{X_{02}}$
$$
\bf{X_{03}}=-\cos{\theta}{\frac{\partial}{\partial{r}}}+\frac{\sin{\theta}}{r}{\frac{\partial}{\partial{\theta}}}
$$
$$
\bf{X_{04}}=-\cos{\phi}\sin{\theta}{\frac{\partial}{\partial{r}}}-\frac{\cos{\phi}\cos{\theta}}{r}{\frac{\partial}{\partial{\theta}}}+
\frac{\sin{\phi}}{r\sin{\theta}}{\frac{\partial}{\partial{\phi}}}~~~~~~~~~~\nonumber
$$
and
$$
\bf{X_{05}}=\sin{\phi}\sin{\theta}{\frac{\partial}{\partial{r}}}+\frac{\sin{\phi}\cos{\theta}}{r}{\frac{\partial}{\partial{\theta}}}
+\frac{\cos{\phi}}{r\sin{\theta}}{\frac{\partial}{\partial{\phi}}}.
$$
The conserved quantities corresponding to $ \bf{X_{03}}$, $
\bf{X_{04}}$ and $ \bf{X_{05}}$ are:
$$
\bf{I_{03}}=2a^2\cos{\theta}{\dot{r}}-2a^2r\sin{\theta}{\dot{\theta}}
$$
$$
\bf{I_{04}}=2a^2\cos{\phi}\sin{\theta}\dot{r}+2a^2r\cos{\phi}\cos{\theta}\dot{\theta}
-2a^2r\sin{\phi}\sin{\theta}\dot{\phi},
$$
$$
\bf{I_{05}}=-2a^2\sin{\phi}\sin{\theta}\dot{r}-2a^2r\sin{\phi}\cos{\theta}\dot{\theta}
-2a^2r\cos{\phi}\sin{\theta}\dot{\phi}.
$$
\begin{table}{Table: $1$}
\begin{center}
\begin{tabular}{|c|c|}
              \hline
  No. & $a(t)$   \\
\hline

              $1$ & $\sin t, \cos t, \tan t$ \\
\hline
              $2$ & $\sinh t, \cosh t, \tanh t$  \\
\hline
              $3$ & $\sqrt {\sin t}, \sqrt {\cos t}, \sqrt {\tan t}$  \\
\hline

              $4$ & $\sqrt {\sinh t}, \sqrt {\cosh t}, \sqrt {\tanh t}$  \\
\hline

              $5$ & $\sqrt {\ln t}$ \\
\hline
              $6$ & $\sqrt{ct+b}$  \\
\hline
              $7$ & $t+\frac{1}{t}$  \\
\hline
              $8$ & $c^2b^2\cosh (\frac{t}{b})^2$  \\
\hline

\end{tabular}
\end{center}
\end{table}
\subsubsection{Six Noether operators}
In this case, we will report some of the forms of cosmic scale
factors which posses six Noether operators.
\\\\($1$): $a(t)=t^n$
\\For $n=2,4,6,...$  we have
 $\bf{X_{01}-X_{05}}$
and
$$
X_{06}=s{\frac{\partial}{\partial{s}}}+\frac{t}{2}{\frac{\partial}{\partial{t}}}-\frac{1}{2}(n-1)r{\frac{\partial}{\partial{r}}}
$$
The associated energy type first integral of motion will be
$$
I_{06}=-s{\dot{t}}^{2}+sa^2{\dot{r}^{2}}+sa^2r^2{\dot{\theta}^{2}}+sa^2r^2\sin^{2}\theta{\dot{\phi}}^{2}+t{\dot{t}}+(n-1)a^2r\dot{r}.
$$
For $n=3,5,7,9,$...and $n= 3+2p$, we have
 $\bf{X_{01}-X_{05}}$
and
$$
X_{06}=s{\frac{\partial}{\partial{s}}}+\frac{t}{2}{\frac{\partial}{\partial{t}}}-(p+1)r{\frac{\partial}{\partial{r}}}.
$$
The corresponding energy type first integral of motion associated
with $ X_{06}$ will be
$$
I_{06}=-s{\dot{t}}^{2}+sa^2{\dot{r}^{2}}+sa^2r^2{\dot{\theta}^{2}}+sa^2r^2\sin^{2}\theta{\dot{\phi}}^{2}+t{\dot{t}}+2(p+1)a^2r\dot{r}.
$$
\\\\($2$): $a(t)=\sqrt{t^n}$
\\For $n=1,3,5,7...$  we have $\bf{X_{01}-X_{05}}$
and
$$
X_{06}=s{\frac{\partial}{\partial{s}}}+\frac{t}{2}{\frac{\partial}{\partial{t}}}-\frac{1}{4}(n-2)r{\frac{\partial}{\partial{r}}}
$$
The associated conserved quantity will be
$$
I_{06}=-s{\dot{t}}^{2}+sa^2{\dot{r}^{2}}+sa^2r^2{\dot{\theta}^{2}}+sa^2r^2\sin^{2}\theta{\dot{\phi}}^{2}+t{\dot{t}}+\frac{1}{2}(n-2)a^2r\dot{r}
$$
For $n=4,8,12,16,....$ we take $n$= $4+4p$ and $p=0,1,2,3....$.\\
The Noether operators are:
 \\$\bf{X_{01}-X_{05}}$
and
$$
X_{06}=s{\frac{\partial}{\partial{s}}}+\frac{t}{2}{\frac{\partial}{\partial{t}}}-\frac{1}{2}(2p+1)r{\frac{\partial}{\partial{r}}}
$$
The corresponding conserved quantity will be
$$
I_{06}=-s{\dot{t}}^{2}+sa^2{\dot{r}^{2}}+sa^2r^2{\dot{\theta}^{2}}+sa^2r^2\sin^{2}\theta{\dot{\phi}}^{2}+t{\dot{t}}+(2p+1)a^2r\dot{r}
$$
For $n=6,10,14,18,20,....$ we take $n$= $6+4p$ and $p=0,1,2,3,....$.
\\The Noether operators are:
 $\bf{X_{01}-X_{05}}$
and
$$
X_{06}=s{\frac{\partial}{\partial{s}}}+\frac{t}{2}{\frac{\partial}{\partial{t}}}-(p+1)r{\frac{\partial}{\partial{r}}}
$$
The associated energy type first integral of motion will be
$$
I_{06}=-s{\dot{t}}^{2}+sa^2{\dot{r}^{2}}+sa^2r^2{\dot{\theta}^{2}}+sa^2r^2\sin^{2}\theta{\dot{\phi}}^{2}+t{\dot{t}}+(p+1)a^2r\dot{r}
$$

\subsubsection{Seven Noether operators}
In this case, we will report some of the forms of cosmic scale
factors which posses seven Noether operators.
\\\\($1$): $a(t)$=$\alpha t+\beta$
\\where $\alpha$ and $\beta$ are arbitrary constant.
\\For this case, we have $\bf{X_{01}-X_{05}}$, the others two are:
\begin{eqnarray}
X_{06}=\frac{s^{2}}{2}{\frac{\partial}{\partial{s}}}+\frac{s}{2}
\bigg(t+\frac{\beta}{\alpha}\bigg){\frac{\partial}{\partial{t}}},
\label{A27}
\end{eqnarray}

\begin{eqnarray}
X_{07}=s{\frac{\partial}{\partial{s}}}+\frac{1}{2}\bigg(t+\frac{\beta}{\alpha}\bigg){\frac{\partial}{\partial{t}}},
\label{A28}
\end{eqnarray}
\begin{eqnarray}
A={\frac{{1}}{2\alpha}}\bigg[(c_{1}t^{2}+2c_{4}) \alpha +2c_{1}\beta
t\bigg].\nonumber
\end{eqnarray}
The associated integral of motions corresponding to  $X_{06}$ and
$X_{07}$ are:
\begin{eqnarray}
I_{06}= -\frac{s^{2}}{2}\dot{t}^{2}+
\frac{s^{2}}{2}a^{2}\dot{r}^{2}+
\frac{s^{2}}{2}a^{2}{r^{4}\dot{\theta}^{2}+\frac{s^{2}}{2}a^{2}r^{4}\dot{\phi}^{2}\sin^{2}\theta}
+st\dot{t}+\frac{s\beta\dot{t}}{\alpha}+\frac{t^{2}}{2}+\frac{\beta}{\alpha},\nonumber
\end{eqnarray}
\begin{eqnarray}
I_{07}=
-s\dot{t}^{2}+sa^{2}\dot{r}^{2}+sa^{2}r^{4}\dot{\theta}^{2}+sa^{2}r^{4}\dot{\phi}^{2}\sin^{2}\theta
+t\dot{t}+\frac{ \beta \dot{t}}{\alpha}. \nonumber
\end{eqnarray}
(2):~$a(t)= \sqrt{t^2}$
\\\\For this case, we have $\bf{X_{01}-X_{05}}$,
\begin{eqnarray}
X_{06}=s{\frac{\partial}{\partial{s}}}+\frac{t}{2}{\frac{\partial}{\partial{t}}},
~~~~~~~~~~X_{07}=\frac{s^{2}}{2}{\frac{\partial}{\partial{s}}}+\frac{st}{2}{\frac{\partial}{\partial{t}}},
~~~~~~~~\label{R1}
\end{eqnarray}
\begin{eqnarray}
A=\frac{c_{1}t^2}{2}+c_{4}.~~~~~~~\nonumber
\end{eqnarray}
The associated conserved quantities will be
\begin{eqnarray}
I_{06}=-s{\dot{t}}^{2}+sa^2{\dot{r}^{2}}+sa^2r^2{\dot{\theta}^{2}}+sa^2r^2\sin^{2}\theta{\dot{\phi}}^{2}+t{\dot{t}},~~~~~~~~~~~~~~\nonumber
\end{eqnarray}
\begin{eqnarray}
I_{07}=-s^2{\dot{t}}^{2}+s^2a^2{\dot{r}^{2}}+
s^2a^2r^2{\dot{\theta}^{2}}+s^2a^2r^2\sin^{2}\theta{\dot{\phi}}^{2}+2st{\dot{t}}-t^2.\nonumber
\end{eqnarray}
\subsubsection{Maximal set of Noether operators}
One can obtain the following set of maximal Noether operators for
$a(t)=$constant i.e. $\bf{X_{01}-X_{03}}$
\begin{eqnarray}
X_{04}=\frac{\partial}{\partial{t}},~X_{05}=\frac{s}{2}{\frac{\partial}{\partial{t}}},
\label{Z1}
\end{eqnarray}

\begin{eqnarray}
X_{06}=(-\frac{s\cos{\theta}}{2}){\frac{\partial}{\partial{r}}}+(\frac{s\sin{\theta}}{2r}){\frac{\partial}{\partial{\theta}}},~~X_{07}=r\cos{\theta}{\frac{\partial}{\partial{t}}}}+t\cos{\theta}{\frac{\partial}{\partial{r}}-(\frac{t\sin{\theta}}{r}){\frac{\partial}{\partial{\theta}}},
\nonumber
\end{eqnarray}

\begin{eqnarray}
X_{08}=s{\frac{\partial}{\partial{s}}}+\frac{t}{2}{\frac{\partial}{\partial{t}}}+\frac{r}{2}{\frac{\partial}{\partial{r}}},~~X_{09}=\frac{s^{2}}{2}{\frac{\partial}{\partial{s}}}+\frac{st}{2}{\frac{\partial}{\partial{t}}}+\frac{sr}{2}{\frac{\partial}{\partial{r}}},
\nonumber
\end{eqnarray}

\begin{eqnarray}
X_{10}=-(\frac{s\sin{\theta}\sin{\phi}}{2}){\frac{\partial}{\partial{r}}}-(\frac{s\cos{\theta}\sin{\phi}}{2r}){\frac{\partial}{\partial{\theta}}}-(\frac{s\cos{\phi}}{2r\sin{\theta}}){\frac{\partial}{\partial{\phi}}}
\nonumber
\end{eqnarray}

\begin{eqnarray}
X_{11}=-(\frac{s\sin{\theta}\cos{\phi}}{2}){\frac{\partial}{\partial{r}}}-(\frac{s\cos{\theta}\cos{\phi}}{2r}){\frac{\partial}{\partial{\theta}}}
+(\frac{s\sin{\phi}}{2r\sin{\theta}}){\frac{\partial}{\partial{\phi}}},
\nonumber
\end{eqnarray}

\begin{eqnarray}
X_{12}=r\sin{\theta}\sin{\phi}{\frac{\partial}{\partial{t}}}+t\sin{\phi}\sin{\theta}{\frac{\partial}{\partial{r}}}
+\frac{t\cos{\theta}\sin{\phi}}{r}{\frac{\partial}{\partial{\theta}}}+(\frac{t\cos{\phi}}{r\sin{\theta}}){\frac{\partial}{\partial{\phi}}},
\nonumber
\end{eqnarray}

\begin{eqnarray}
X_{13}=r\sin{\theta}\cos{\phi}{\frac{\partial}{\partial{t}}}+t\cos{\phi}\sin{\theta}{\frac{\partial}{\partial{r}}}\nonumber
+\frac{t\cos{\theta}\cos{\phi}}{r}{\frac{\partial}{\partial{\theta}}}-(\frac{t\sin{\phi}}{r\sin{\theta}}){\frac{\partial}{\partial{\phi}}}.
\nonumber
\end{eqnarray}
\begin{eqnarray}
\nonumber
\end{eqnarray}

The corresponding energy type first integral of motions associated
with ${X_{04}-X_{13}}$ are
\begin{eqnarray}
I_{04}=2{\dot{t}}, ~~~~I_{05}=s{\dot{t}}-t,\nonumber
\end{eqnarray}
\begin{eqnarray}
I_{06}=\frac{sa^2\cos{\theta}{\dot{r}}}{2}-sa^2r^2\sin{\theta}{\dot{\theta}}-r\cos{\theta},\nonumber
\end{eqnarray}
\begin{eqnarray}
I_{07}=2r\cos{\theta}{\dot{t}}-2a^2t\cos{\theta}{\dot{r}}-2a^2rt\sin{\theta}{\dot{\theta}},
 \nonumber
\end{eqnarray}
\begin{eqnarray}
I_{08}=-s{\dot{t}}^{2}+sa^2{\dot{r}^{2}}+sa^2r^2{\dot{\theta}^{2}}+sa^2r^2\sin^{2}\theta{\dot{\phi}}^{2}+t{\dot{t}}-ra^2{\dot{r}},\nonumber
\end{eqnarray}
\begin{eqnarray}
I_{09}=-\frac{s^2{\dot{t}}^{2}}{2}+\frac{s^2a^2{\dot{r}^{2}}}{2}+
\frac{s^2a^2r^2{\dot{\theta}^{2}}}{2}+\frac{s^2a^2r^2\sin^{2}\theta{\dot{\phi}}^{2}}{2}+st{\dot{t}}-sra^2{\dot{r}}-\frac{1}{2}(t^2-r^2),\nonumber
\end{eqnarray}
\begin{eqnarray}
I_{10}=a^{2}s\sin{\theta}\sin{\phi}\dot{r}+a^{2}rs\sin{\phi}\cos{\theta}\dot{\theta}
+sra^2\cos{\phi}\sin{\theta}\dot{\phi}-r\sin{\theta}\sin{\phi},\nonumber
\end{eqnarray}
\begin{eqnarray}
I_{11}=a^{2}s\sin{\theta}\cos{\phi}\dot{r}+a^{2}rs\cos{\phi}\cos{\theta}\dot{\theta}-sra^2\sin{\phi}\sin{\theta}\dot{\phi}-r\sin{\theta}\cos{\phi},\nonumber
\end{eqnarray}
\begin{eqnarray}
I_{12}=2r\sin{\theta}\sin{\phi}{\dot{t}}-2ta^{2}\sin{\theta}\sin{\phi}\dot{r}-2tra^{2}\cos{\theta}\sin{\phi}\dot{\theta}-2a^2rt\cos{\phi}}{\sin{\theta}\dot{\phi},\nonumber
\end{eqnarray}
\begin{eqnarray}
I_{13}=2r\sin{\theta}\cos{\phi}{\dot{t}}-2ta^{2}\sin{\theta}\cos{\phi}\dot{r}-2tra^{2}\cos{\theta}\cos{\phi}\dot{\theta}
-2a^2rt\sin{\phi}}{\sin{\theta}\dot{\phi}.\nonumber
\end{eqnarray}

\subsection{For open universe ($k= 1$)}
Using curvature $k= 1$ in Eq.~(\ref{c6}) the Lagrangian becomes
\begin{eqnarray}
L=
{\dot{t}}^{2}-\bigg(\frac{a^2{\dot{r}}^{2}}{1-r^2}\bigg)-a^2r^2{\dot{\theta}}^{2}-a^2r^2{\dot{\phi}}^{2}\sin^{2}\theta.
 \label{c17}
 \end{eqnarray}
 Using Eq.~(\ref{c17}) in Eq.~(\ref{T4}) and after some tedious calculation we get an over-determined system of linear PDEs.

\subsubsection{Minimal set of Noether operators}
 For arbitrary choice of $a(t)$ one can obtain minimal number of Noether
 symmetries i.e. $\bf{X_{01}-X_{02}}$.

\subsubsection{Three Noether operators}
Some different forms of $a(t)$ which posses three Noether operators
are listed in Table: $1$ and the set of Noether symmetries consist
of $\bf{X_{01}-X_{02}}$ and
\begin{eqnarray}
X_{03}=(-\sqrt{r-1}\sqrt{r+1}\cos\theta){\frac{\partial}{\partial{r}}}+(\frac{\sqrt{r-1}\sqrt{r+1}\sin\theta}{r}){\frac{\partial}{\partial{\theta}}},~~~
\label{A17}
\end{eqnarray}
The corresponding integral of motion for $X_{08}$ is:
\begin{eqnarray}
I_{03}=
\frac{2a^{2}{\dot{r}}}{1-r^{2}}(\sqrt{r-1}\sqrt{r+1}\cos\theta)-2a^2r{\dot{\theta}}(\sqrt{r-1}\sqrt{r+1}\sin\theta).
 \nonumber
\end{eqnarray}

\subsubsection{Five Noether operators}
In this case, we will report some of the forms of cosmic scale
factors which posses five Noether operators.
\\\\($1$): $a(t)$=$\alpha t+\beta$
\\where $\alpha$ and $\beta$ are arbitrary constant.
For this case, we have $\bf{X_{01}-X_{02}}$, (\ref{A27}),
(\ref{A28}) and (\ref{A17}) with gauge function:
\begin{eqnarray}
A={\frac{{1}}{2\alpha}}\bigg[(c_{1}t^{2}+2c_{4}) \alpha +2c_{1}\beta
t\bigg].
\end{eqnarray}

\subsubsection{Maximal set of Noether operators}
One can obtain the following set of maximal Noether operators for
$a(t)=$constant i.e.
\\$\bf{X_{01}-X_{02}}$,(\ref{Z1}) and (\ref{A17})
where gauge function is $A= c_{2}t+ c_{3}$.

\subsection{For closed universe ($k= -1$)}
Using curvature $k= -1$ in Eq.~(\ref{c6}) the Lagrangian becomes
\begin{eqnarray}
L=
{\dot{t}}^{2}-\bigg(\frac{a^2{\dot{r}}^{2}}{1+r^2}\bigg)-a^2r^2{\dot{\theta}}^{2}-a^2r^2{\dot{\phi}}^{2}\sin^{2}\theta.
 \label{c27}
 \end{eqnarray}
 Using Eq.~(\ref{c27}) in Eq.~(\ref{T4}) and after some tedious calculation we get an over-determined system of linear PDEs.

\subsubsection{Minimal set of Noether operators}
 If $a(t)$ is arbitrary then we get minimal number of Noether
 symmetries i.e. $\bf{X_{01}-X_{02}}$.

\subsubsection{Five Noether operators}
Some different forms of $a(t)$ which are listed in Table: $1$ posses
five Noether operators i.e. $\bf{X_{01}-X_{02}}$ and others

\begin{eqnarray}
X_{03}=(-\sqrt{1+r^2}\cos\phi\sin\theta){\frac{\partial}{\partial{r}}}-(\frac{\sqrt{1+r^2}\cos\theta\cos\phi}{r}){\frac{\partial}{\partial{\theta}}}+(\frac{\sqrt{1+r^2}\sin\phi}{r\sin\theta}){\frac{\partial}{\partial{\phi}}},~~~~~~
 \label{p27}
\end{eqnarray}
\begin{eqnarray}
X_{04}=(\sqrt{1+r^2}\sin\phi\sin\theta){\frac{\partial}{\partial{r}}}+(\frac{\sqrt{1+r^2}\cos\theta\sin\phi}{r}){\frac{\partial}{\partial{\theta}}}+(\frac{\sqrt{1+r^2}\cos\phi}{r\sin\theta}){\frac{\partial}{\partial{\phi}}},~~~~~~~~
 \label{p28}
\end{eqnarray}
\begin{eqnarray}
X_{05}=(-\sqrt{1+r^2}\cos\theta){\frac{\partial}{\partial{r}}}+(\frac{\sqrt{1+r^2}\sin\theta}{r}){\frac{\partial}{\partial{\theta}}}.~~~~~~~~~~~~~~~~
 \label{p29}
\end{eqnarray}
The energy type first integral of motions associated with above
generators are
\begin{eqnarray}
I_{03}=  \frac{2a^2\dot{r}}{\sqrt{1+r^2}}\cos\phi\sin\theta
+2a^2r\sqrt{1+r^2}\cos\theta\cos\phi\dot{\theta}-2a^2r\sqrt{1+r^2}\sin\theta\sin\phi\dot{\phi},
\nonumber
\end{eqnarray}
\begin{eqnarray}
I_{04}=  -\frac{2a^2\dot{r}}{\sqrt{1+r^2}}\sin\phi\sin\theta
-2a^2r\sqrt{1+r^2}\cos\theta\sin\phi\dot{\theta}-2a^2r\sqrt{1+r^2}\sin\theta\cos\phi\dot{\phi},
\nonumber
\end{eqnarray}
\begin{eqnarray}
I_{05}=  \frac{2a^2\dot{r}}{\sqrt{1+r^2}}\cos\theta
-2a^2r\sqrt{1+r^2}\sin\theta\dot{\theta}.~~~~~~~~~~~~~~~~~~~~~~~~~~~~~~~~~~~~~~~~~
 \nonumber
\end{eqnarray}

\subsubsection{Maximal set of Noether operators}
($1$): $a(t)$=$\alpha t+\beta$
\\where $\alpha$ and $\beta$ are arbitrary constant.
For this the Noether operators are:
\\$\bf{X_{01}-X_{02}}$, (\ref{A27}), (\ref{A28}) and (\ref{p27})-(\ref{p29}) with the gauge
function:
\begin{eqnarray}
A={\frac{{1}}{2\alpha}}\bigg[(c_{1}t^2+2c_{4})\alpha +2c_{1}\beta
t\bigg].~~~~~~~~~~~~~~~~~~~~~~~~~~~~~~~
\end{eqnarray}
\\($2$): If $a(t)$ is constant then we have
\\ $\bf{X_{01}-X_{02}}$, (\ref{Z1}) and
 (\ref{p27})-(\ref{p29}), where
\begin{eqnarray}
A=c_{2}t+c_{3}.~~~~~\nonumber
\end{eqnarray}
It should be noted that  $\bf{X_{01}-X_{02}}$ not only independent
from the choice of cosmic scale factor but also from the values of
$k$.

\section{Conclusion}

In this paper, a classification of scale factor by considering
different forms of FRW universe i.e. flat, open and closed was
reported by using Noether approach. The Noether symmetries in the
presence of gauge term was computed. We distributed our results with
respect to different natures of universe. For the classification
purpose, and an unusual Lagrangian for FRW metric was considered.
With the help of considered Lagrangian and Noether equation, an
over-determined system of PDEs was obtained. Then these systems were
solved for different values of cosmic scale factor $a(t)$. Cases
were reported with respect to different number of Noether operators.
For all three natures of universe minimal and maximal set of Noether
operators were computed. For each calculated Noether operator
corresponding energy type first integral of motion was presented. It
was reported that $\bf{X_{01}-X_{02}}$ not only independent from the
choice of cosmic scale factor but also the natures of universe.
{}
\end{document}